\documentclass[journal]{IEEEtran}
\usepackage{amsmath,amssymb}
\usepackage{subfigure}
\pdfoutput=1
\usepackage{graphicx,graphics,color,psfrag}
\usepackage{cite,balance}
\usepackage{caption}
\allowdisplaybreaks
\usepackage{algorithm}
\usepackage{algorithmic}
\usepackage{accents}
\usepackage{amsthm}
\usepackage{bm}
\usepackage{url}
\usepackage[english]{babel}
\usepackage{multirow}
\usepackage{enumerate}
\usepackage{cases}
\usepackage{stfloats}
\usepackage{dsfont}
\usepackage{color,soul}
\usepackage{amsfonts}
\usepackage{cite,graphicx,amsmath,amssymb}
\usepackage{subfigure}
\usepackage{fancyhdr}
\usepackage{hhline}
\usepackage{graphicx,graphics}
\usepackage{array,color}
\usepackage{mathtools}
\usepackage{amsmath}

\include{header}
\begin{document}
\captionsetup[figure]{name={Figure}}
%\title{\huge 
%%\vspace{-10pt}
%%\title{\LARGE 
%UAV-Enabled Data Harvesting in Probabilistic LoS Channels}
%\author{Changsheng You and Rui Zhang   \thanks{\noindent C. You and R. Zhang are  with the Dept. of Electrical and Computer Engineering, National University of Singapore, Singapore (Email: eleyouc@nus.edu.sg, elezhang@nus.edu.sg).
% }}

\title{\huge 
How to Deploy Intelligent Reflecting Surfaces in Wireless Network:  BS-side, User-side, or Both Sides?} \vspace{-3pt}
\author{Changsheng You, Beixiong Zheng, Weidong Mei,
	 and Rui Zhang 
	   \thanks{\noindent    
	   C. You is with the Department of Electronic and Electrical Engineering, Southern University of Science and Technology (SUSTech), Shenzhen 518055, China (e-mail: youcs@sustech.edu.cn). He was with the Department of Electrical and Computer Engineering, National University of Singapore, Singapore 117583, Singapore.
	    B. Zheng, W. Mei, and R. Zhang are with the Department of Electrical and Computer Engineering, National University of Singapore, Singapore 117583, Singapore (e-mail: {elezbe, wmei, elezhang}@nus.edu.sg).
%	   The authors are with the Department of Electrical and Computer Engineering, National University of Singapore 117583, Singapore (Email: \{eleyouc, elezbe, elezhang\}@nus.edu.sg).
}}  
\maketitle

\begin{abstract} 
The performance of wireless communication systems is fundamentally constrained by the random and uncontrollable wireless channel. By leveraging the recent advances in digitally-controlled metasurface, intelligent reflecting surface (IRS) has emerged as a promising solution to enhance the wireless network performance by smartly reconfiguring  the radio propagation environment. Despite the substantial research on IRS-aided communications, this article addresses the important issue of how to deploy IRSs in a wireless network to achieve its optimum performance. We first compare the two conventional strategies of deploying IRS at the side of base station or  users in terms of various communication performance metrics, and then propose a new \emph{hybrid} IRS deployment strategy  by combining their complementary advantages. Moreover, the main challenges in optimizing IRS deployment as well as  their promising solutions are discussed.  Numerical results are also presented to compare the performance of different IRS deployment strategies and draw useful insights for practical design.
\end{abstract} 

\begin{IEEEkeywords}Intelligent reflecting surface (IRS), IRS deployment, Double-IRS system, Cooperative beamforming, Channel estimation
\end{IEEEkeywords}

%\footnote{This work is supported by Ministry of Education, Singapore under Award T2EP50120-0024 and by Advanced Research and Technology Innovation Centre (ARTIC) of National University of Singapore under Research Grant R-261-518-005-720. (\emph{Corresponding author: Beixiong Zheng.})\\\indent
%C. You is with the Department of Electronic and Electrical Engineering, Southern University of Science and Technology (SUSTech), Shenzhen 518055, China (e-mail: youcs@sustech.edu.cn). He was with the Department of Electrical and Computer Engineering, National University of Singapore, Singapore 117583, Singapore.\\
%\indent B. Zheng, W. Mei, and R. Zhang are with the Department of Electrical and Computer Engineering, National University of Singapore, Singapore 117583, Singapore (e-mail: {elezbe, wmei, elezhang}@nus.edu.sg).\\
%%Firstname1 Lastname1, Firstname2 Lastname2. Address1, City1 zipcode1, Country1 (e-mail: email1; email2).\\\indent
%%Firstname1 Lastname1. Address2, City2 zipcode2, Country2 (e-mail: email1).\\\indent
%%Firstname2 Lastname2. Address3, City3 zipcode3, Country3 (e-mail: email2).
%}

%%%%%%%%%%%%%%%%%%%%  main text   %%%%%%%%%%%%%%%%%%%%%%%%%%%%%%%%

\section{Introduction}

While the fifth-generation (5G) wireless network is under deployment globally, researchers have been enthusiastically fostering the future  sixth-generation (6G) wireless network that targets at supporting new and promising Internet-of-everything (IoE) applications ranging from extended reality, automated system, to tactile Internet. These applications impose more stringent performance requirements than 5G, such as ultra-high data rate, global coverage and connectivity, and extremely high reliability and low latency, which may not be fully achieved by existing  technologies for 5G.

%To enable the 6G applications, 
To meet the future demands of 6G, an innovative concept of \emph{smart radio environment} has been recently proposed\cite{Renzo2019Smart,qingqing2019towards}. Specifically, the radio prorogation environment, traditionally deemed to be random and largely uncontrollable, can be dynamically reconfigured  to enhance the wireless communication performance,  by leveraging the digitally-controlled passive metasurface\cite{Renzo2019Smart,qingqing2019towards}.  This promising application of metasurface to wireless communications has rapidly spurred intensive research into the new technology of   \emph{intelligent reflecting surface} (IRS)\cite{qingqing2019towards} or its various equivalents such as  reconfigurable intelligent surface (RIS)\cite{basar2019wireless}.
% and others\cite{Renzo2019Smart,liang2019large}.
  The current research on IRS/RIS can be roughly classified into two main categories, namely, hardware design and system modeling\cite{di2020smart}, and communication design and performance study\cite{wu2020intelligent}. {\color{black}Specifically, 
IRS can be  utilized to realize a variety of key functions in wireless communication systems by dynamically controlling the amplitudes and/or phases of its reflected signals via its smart controller},  such as  creating (virtual) line-of-sight (LoS) links, improving the channel rank condition,   
%bypassing environment obstacles between transceivers, improving channel rank condition, 
enhancing desired signal power and/or suppressing co-channel interference by passive beamforming, reshaping the  channel realization and/or statistical distribution, and so on\cite{wu2020intelligent,huang2018largeRIS}. {\color{black}Moreover, IRS operates in full-duplex (FD) mode with passive reflection only and thus is free of amplification/processing noise as well as self-interference.}
These appealing functions of IRS have motivated substantial studies on applying it to  boost the wireless system performance in various scenarios, such as multi-antenna/multi-carrier communications, multi-user   
%attracted upsurge attention in recent years on the IRS designs 
%%designing new IRS-aided wireless networks 
%with various wireless technologies, such as
 non-orthogonal multiple access (NOMA), physical-layer security, mobile edge computing (MEC), etc\cite{wu2020intelligent,liu2020reconfigurable}.

\begin{figure}[t]
\begin{center}
\includegraphics[width=8.7cm]{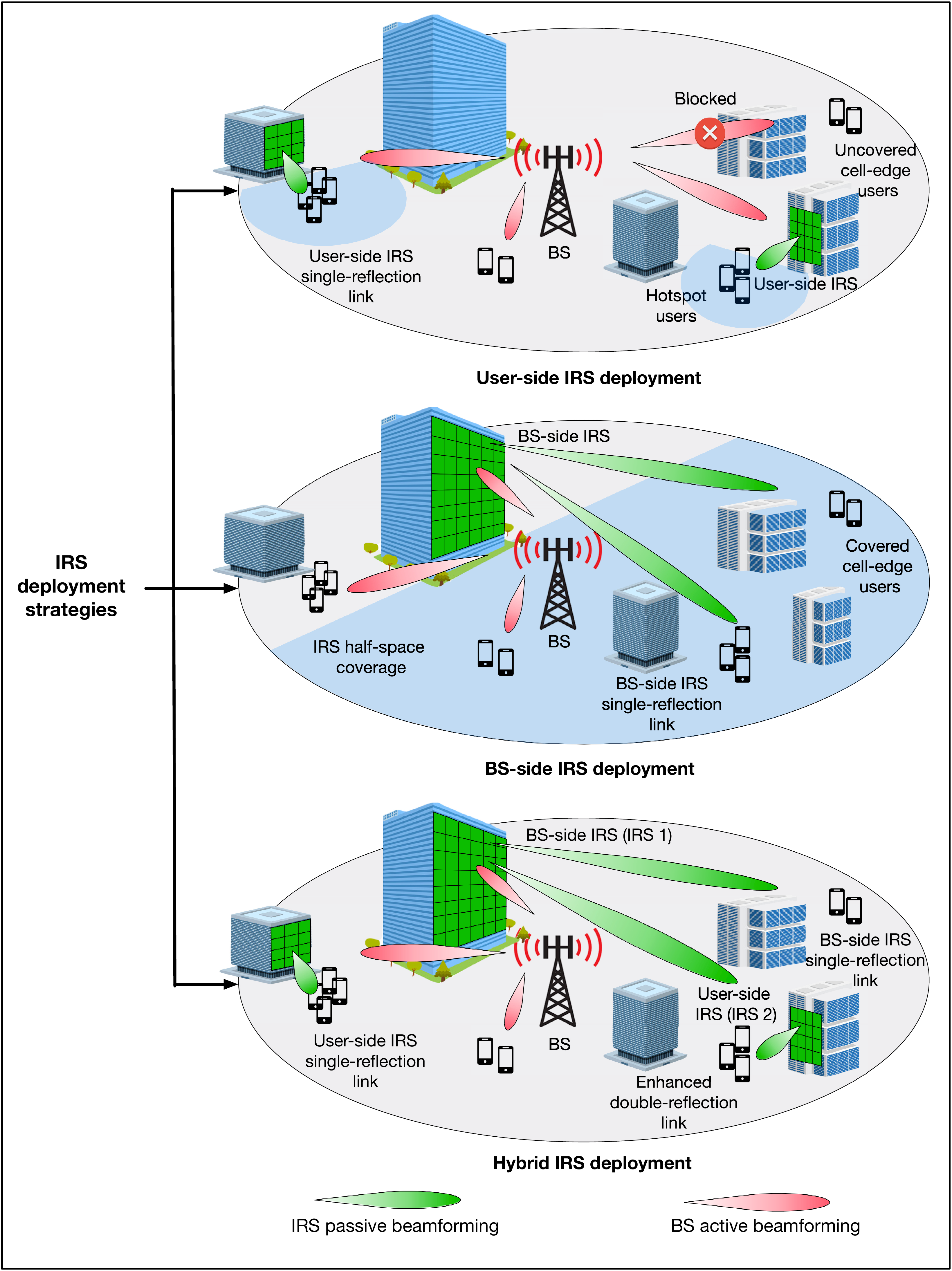}
\caption{Illustration of different IRS deployment strategies.
}
\label{Fig:deployment}
\end{center}
\end{figure}

In the existing works on IRS, it is usually deployed at the side of distributed users in the network to help enhance the communication performance with their  aided base stations (BSs),    
 as illustrated in the upper part of  Fig.~\ref{Fig:deployment}\cite{dunna2020scattermimo}. In practice, this {\it user-side IRS deployment} is preferred at designated locations/places such as  
   hotspot, cell edge, and moving vehicle for enhancing the local coverage. It is worth noting that this deployment strategy is fundamentally due to the   passive signal reflection of IRS  for minimizing its severe  
   product-distance path-loss over the two links with its associated remote BS and local user, respectively.  Thus, this deployment is  drastically  different from that for the conventional 
    active relay which can amplify  the signal from the source before forwarding it to the destination, and hence      
     is usually placed in the middle of them to achieve the best performance.
 Alternatively, the other IRS deployment  strategy that is also able  to minimize  the product-distance path-loss is 
% by deploying IRS near the base station (BS), referred to as
  the \emph{BS-side IRS deployment}, shown in the middle part of  Fig.~\ref{Fig:deployment}. 
  Note that this IRS deployment resembles that for the conventional reflect-array
%  which is also placed near the transmitter 
  in e.g., satellite communications; while they generally differ in distance with the source (far- versus near-field) as well as main function (smart channel reconfiguration versus low-cost beamforming).   In the following, we compare the user- versus  BS-side IRS deployment strategies in terms of their key metrics including network coverage, reconfigured channel condition, passive beamforming performance, and signaling overhead.
\begin{itemize}
\item[1)] {\bf Network coverage:} 
%{\color{blue}
As user-side IRSs serve users in their vicinities only, they are unable to assist users at other locations in the network, especially when the  user distributions are unknown and random in the network. In contrast, the BS-side IRS has the potential to reconfigure channels for all users in the network, thus generally offering a larger network coverage than user-side IRSs. 
This is fundamentally due to the smaller path-loss in the BS-IRS link with the BS-side IRS deployment as well as the squared power scaling order of IRS  that  leads to a much higher passive beamforming gain at the (centralized) BS-side IRS than multiple smaller-size (distributed) user-side IRSs.
Moreover, it is worth noting that for IRSs coated on objects such as walls and facades of buildings, they  can serve users in \emph{half of the space} only (see Fig.~\ref{Fig:deployment}), for both user-/BS-side IRS  deployment strategies.
\item[2)] {\bf Reconfigured channel condition:} 
%{\color{blue}
One key function of IRS is  to bypass environmental obstacles for creating virtual  LoS  BS-user  links via IRS smart reflection. To achieve this goal, IRS can be properly deployed at the user (or BS)  side to establish LoS links with its nearby users (or BS) with high probability owing to their short distances. Besides, the user-side IRS has an additional advantage that, with proper placement, it is more likely to establish  a virtual LoS(-dominant)  link with the BS;
% establish  with proper placement, it can establish an LoS link with the BS with high probability, 
 whereas it is generally difficult  for a single  BS-side IRS to enjoy  LoS links with all users in the network due to their random locations. Generally speaking, BS-side or user-side IRSs alone may not be able to guarantee achieving obstacle-free  links between the BS and all users in the network, as shown in Fig.~\ref{Fig:deployment}.     
\item[3)] {\bf Passive beamforming performance:}
The BS-side IRS  in general outperforms its user-side counterparts given the same budget on the  total number of reflecting elements  in terms of the maximum passive beamforming gain achievable for each user. {\color{black}This is because each user can be served by
all reflecting elements at the BS-side IRS,  while it obtains a  smaller passive beamforming gain from its nearby user-side  IRSs only  since the signals reflected by other far-apart user-side IRSs are much weaker due to the higher path-loss\cite{qingqing2019towards}.}
Moreover, the passive beamforming at the BS-side IRS can be more flexibly  adjusted according to all users'  channels and different quality-of-service (QoS) requirements as compared to user-side IRSs that can only cater to their locally served users.
%, thus providing more flexibility than that of the user-side IRS.
\item[4)] {\color{black}{\bf IRS-BS signaling overhead:} 
The IRS reflection design in general requires information exchange between the IRS controller  and its associated BS to tune IRS reflection coefficients. As compared to user-side IRSs, the BS-side IRS generally requires lower signaling overhead with the associated BS, thanks to its much shorter distance  with the BS and hence higher signaling rate, as well as dispensing with the need of coordination among multiple distributed IRSs.}
\end{itemize}

 \begin{table*}[t!]
\centering
{
\caption{Comparison of three IRS deployment strategies.}\label{Table:Comp}
\resizebox{1\textwidth}{!}{
\begin{tabular}{| c | c |c|c|c|c|}
%\begin{tabular}{| m{3cm} | m{1.4cm} |m{1.4cm}|m{2.6cm}|m{1.2cm}|m{1.2cm}|}
\hline
\multirow{2}{*}{\textbf{IRS deployment strategy}}&\multirow{1}{*}{\textbf{Network}}&\multirow{1}{*}{\textbf{Reconfigured channel }}&\multirow{1}{*}{\textbf{Passive beamforming}}&\multirow{1}{*}{\textbf{IRS-BS signaling}}&\multirow{1}{*}{\textbf{Double-reflection}}\\
%\textbf{strategy}
& \textbf{coverage}& \textbf{LoS availability}& \textbf{gain per user}& \textbf{overhead} &\textbf{link }\\
%\hline
%\textbf{Deployment strategy}& \textbf{Network coverage}& \textbf{Channel condition}&\textbf{Passive beamforming performance} \\
\hline
User-side deployment & Relatively small & High& Relatively small&High & No \\
\hline
BS-side deployment & Large &  Relatively low& Large & Low& No \\
\hline
Hybrid deployment & Very large & Very high& Very large & High& Yes\\
\hline
\end{tabular}}
}
\end{table*}

Motived by the above, we further propose a more general 
 \emph{hybrid IRS deployment} strategy by combining the complementary advantages of the BS- and user-side IRSs with more flexibility to trade-off between them, as shown in the lower part of Fig.~\ref{Fig:deployment}.  Moreover,  besides the  IRS single-reflection (i.e., BS-IRS-user) links in the user-/BS-side deployment cases, 
%each corresponding to the IRS near the BS/users,
the proposed hybrid IRS deployment leads to an additional inter-IRS reflection link between the BS-side IRS and each of the user-side IRSs (e.g., see the  BS-IRS $1$-IRS ${2}$-user link shown in Fig.~\ref{Fig:deployment}).
The \emph{double-reflection} links can be exploited to establish more available LoS paths between the BS and its served users, especially when  both the direct and single-reflection links between them  are severely  blocked.
% (e.g., the communications along the road corners). 
 More interesting, it has been shown in\cite{Han2020Cooperative} that under the LoS channel setup, 
 the double-reflection link can achieve a much higher asymptotic passive beamforming gain as compared to the single-reflection link given the same total number of reflecting elements, denoted by $N$, as $N$ becomes sufficiently large (i.e., in the order of $\mathcal{O}({N^4})$ versus $\mathcal{O}(N^2)$), despite the fact that the former suffers more path-loss than the latter due to the double (versus single) signal reflection. This have motivated substantial research interesting recently on investigating the design of double-IRS aided wireless systems\cite{Han2020Cooperative,you2020wireless,yildirim2020modeling,shao2021joint,abdullah2021double}.
 
Although  the hybrid IRS deployment potentially offers better flexibility and hence superior   performance as compared to the user-/BS-side IRS deployment alone, it also incurs higher complexity in design and implementation, explained as follows.  
% generally requires higher complexity and implementation cost than the BS/user-side IRS deployment, as well as faces new design challenges.
First, how to optimally assign the reflecting elements at the BS-side IRS to serve users directly (i.e., via the BS-side single-reflection links only) or with some of the user-side IRSs cooperatively (i.e., via the double-reflection links over them) is an important but difficult problem to solve. This problem is further complicated with the joint design of the  active beamforming of the BS and the passive beamforming of all IRSs.   
%ly, the IRS association and users' (reflection) mode selection (among the double-reflection and user/BS-side IRS single-reflection modes) need to be carefully designed to balance the communication performance among different users. 
Second, 
%under the complex radio propagation environment,
 how to efficiently acquire the necessary channel state information (CSI) for the above-mentioned IRS association and joint active/passive beamforming  design is more  practically   challenging than the case with user-/BS-side IRSs only,     due to the presence of both the single- and double-reflection links and their intricate coupling. Third, both the BS- and user-side IRSs need to be properly placed in the network, and the reflecting elements allocation among them should be optimized for maximally enhancing the network performance, which is also a hard problem to solve in practice.  In  Table~\ref{Table:Comp}, we compare the three IRS deployment strategies  in various main  aspects.

\section{Design Challenges and Promising Solutions }
 In this section, we present the main challenges in designing the hybrid IRS deployment, including the IRS-user  association and mode selection, CSI acquisition and beamforming optimization, as well as  IRS placement and elements allocation. Promising approaches are also proposed to solve them.

\subsection{{IRS-User  Association and Mode Selection}}

\begin{figure*}[t!]
%\vspace{5pt}
\begin{center}
\includegraphics[width=16cm]{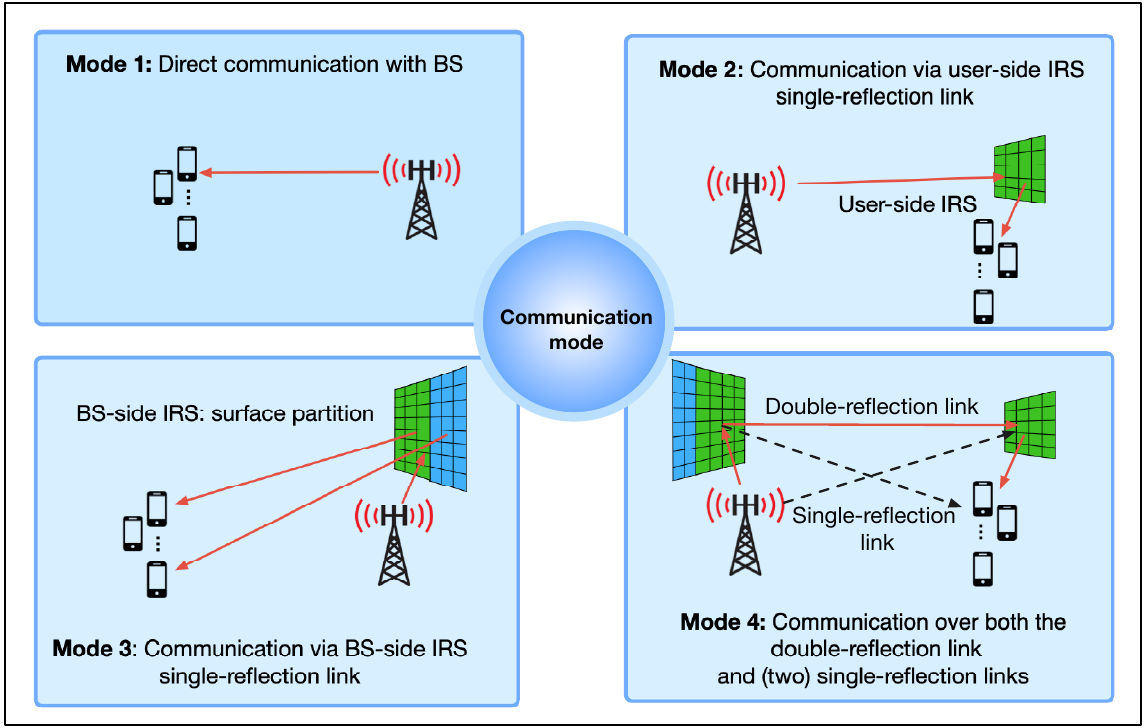}
\caption{Illustration of different  communication modes in an IRS-aided system.}
\label{Fig:mode}
\end{center}
\end{figure*}

Under the hybrid IRS deployment, the foremost challenge in optimizing the wireless network performance is the design of  IRS-user  association. {\color{black}Compared to the user-/BS-side IRS deployment, IRS-user  association under the hybrid IRS deployment is  more involved. In this case, each of the users may be associated with no IRS, the BS-side IRS only, some user-side IRSs only, or both the BS- and user-side IRSs.} The specific association design depends on the user's communication mode, which can be direct communication with the BS only, communication with the BS over the user-/BS-side IRS single-reflection link only,  or over both the (two) single-reflection and (one) double-reflection links together, as illustrated  in Fig.~\ref{Fig:mode}. 
Generally speaking, the design of IRS-user  association and (user) mode selection (IA-MS) needs to take into account various practical factors 
 % may be affected by many practical factors, 
 such as users' QoS requirements, channel conditions, and co-channel interference levels. Besides, the IA-MS design is also coupled with  
 the active beamforming design at the BS and other communication resources (e.g., power, bandwidth) allocation,  users' multiple access scheme and transmission scheduling, as well as the reflection coefficients at all IRSs in the network. 
 %  the joint (active and passive) beamforming  since different beamforming schemes may result in different optimal IA-MS schemes, and vice versa. 
  Moreover, as the users move within the same cell, their associated BS may be unchanged, while their associated IRSs and communication modes usually  need to be adaptively adjusted, especially for high-mobility users without any user-side IRS attached.

However, the optimal IA-MS design is practically challenging and may be even impossible. This is because 1) it requires the full  CSI of all involved communication links in the network which is difficult to obtain in practice; and 2) it is intricately coupled with the active/passive  beamforming design and hence cannot be solved alone, even with perfect CSI.  {{\color{black}To tackle the above difficulties, a practically appealing approach is to   leverage the \emph{statistical CSI} for designing the 
%of some dominant links to design the statistically-favorable
 IA-MS based on a practical  \emph{surface-partition} scheme. Specifically, 
for the  BS-side IRS, the statistical CSI (e.g., the deterministic large-scale fading channel components  and fading channel spatial correlation matrix) of its links with the associated BS and other user-side IRSs 
(as well as their locally served users due to close proximity)  
  can be first obtained (by e.g., equipping the IRS with low-cost sensors). Moreover, the BS-side IRS can be partitioned into multiple subsurfaces, which consist of equal or unequal number of reflecting elements (see Modes 3 and 4 in  Fig.~\ref{Fig:mode}). As such,       
  instead of associating multiple  users to all elements of each  BS-side IRS 
  at the same time for joint design,
a low-complexity scheme is to assign each subsurface to one  user only at each time, based on the statistical CSI, users' QoS requirements, and their estimated link signal-to-noise ratios (SNRs) with/without a helping user-side IRS. For example, under the binary LoS/non-LoS (NLoS) channel model and based on the large-scale channel gains (e.g., path loss) of all involved links, the received signal power of each user's link with respect to the number of IRS subsurfaces assigned  can be estimated. Accordingly, a simple  integer programming can be formulated to optimize the IRS surface-partition for optimizing the performance, e.g., by  maximizing the minimum received signal power among all users. In this case, the user with a smaller channel power gain should be allocated with more IRS subsurfaces.
While the surface-partition scheme also applies to each of the user-side IRSs, its surface size is practically much smaller than that of the BS-side IRS, thus rendering its channels with locally served users more likely to be spatially  correlated. Therefore, instead of applying the surface-partition scheme,  a more efficient IRS-user association method for the user-side IRS is to leverage its time-varying reflections to serve its locally associated users over different time\cite{mei2020performance}.} Accordingly, the SNRs of these users can be estimated and then fed back to the BS for facilitating the design of the BS-side IRS association as well as determining whether there is a need for each user to be associated with its nearest user-side IRS, if any.       

It is worth noting that although the proposed IA-MS design based on statistical CSI and surface partition  is suboptimal in general, it significantly reduces the design complexity and also helps  simplify the subsequent CSI acquisition and  beamforming design  (as will be detailed in the next subsection).
 Moreover, the proposed IA-MS design can be conveniently adjusted over time for different users  according to their movement and/or sporadic traffic by simply re-assigning each IRS subsurface to a different user, without affecting other subsurfaces and their user associations.    
Despite the above advantages, the proposed  IA-MS design also faces several new challenges. 
 First, the performance loss of the suboptimal~surface-partition approach needs to be characterized as compared to the optimal IRS-user associations without surface partition.     
 Next, the IA-MS design based on statistical CSI can be further improved/refined 
   based on the actual SNRs of the users.  Efficient online algorithms such as that based on  Kalman filter can be applied to track users' SNRs and adaptively adjust the IRSs' passive beamforming.

\subsection{CSI Acquisition and Beamforming Optimization}

With the proposed  IA-MS design, all users' IRS associations and communication modes are determined. Then, efficient CSI acquisition and joint active/passive  beamforming designs  can be implemented  for optimizing the users' QoS performance,  which can be roughly classified into two categories, namely, \emph{channel-estimation based} and \emph{beam-training based} methods.

The channel-estimation based method needs to firstly estimate the CSI of all involved  links and then design the joint active/passive  beamforming based on estimated CSI. For IRS channel estimation under the user-side IRS deployment, there are two main approaches in the existing literature based on different IRS configurations.
% depending on whether sensing devices  are integrated into IRS or not. 
 First, for semi-passive IRS mounted with sensing devices for receiving signals, {\color{black}the CSI of  the BS/user$\to$IRS links  can be constructed from that of the BS/user$\to$sensor links by using e.g., data-interpolation and  channel-calibration techniques\cite{taha2021enabling}.} 
 However, this method only works for time-division duplexing (TDD) systems based on channel reciprocity, while it is infeasible for frequency-division duplexing (FDD) systems.
Second, for (fully) passive IRS without sensing devices integrated, a viable channel estimation approach is by estimating the cascaded BS-IRS-user reflection channel at the BS/users based on uplink/downlink pilot signals, respectively, while the IRS should dynamically tune its training reflection over time to facilitate the cascaded channel estimation\cite{wu2020intelligent,jensen2019optimal,he2019cascaded}. {\color{black}However, when extending this method to the general case with the hybrid IRS deployment, a new challenge 
arises due to  the double-reflection link since its cascaded CSI is more difficult to estimate.
  This is because not only more channel coefficients are involved in the double-reflection link, but also the double-reflection link is intricately coupled with the two single-reflection links (see Mode 4 in Fig.~\ref{Fig:mode}). To address this problem, efficient cascaded channel estimation schemes have been recently proposed in\cite{you2020wireless} and\cite{zheng2020efficient} to estimate the double-reflection channel for the single-user case, and both single- and double-reflection channels for a cluster of users, respectively.
These results can be further extended to the case with arbitrary channel condition and user location under the hybrid IRS deployment.}
%  still needs further investigation. 
{\color{black}On the other hand, it is also interesting to study how to apply deep learning techniques to improve the channel estimation performance under the hybrid IRS deployment\cite{liu2020deep}.}
%. For semi-passive IRS, one can recover the inter-IRS channel (i.e., IRS $1$-IRS $2$) from the CSI of the IRS $1$-IRS $2$-user and IRS $2$-user links estimated.  
%However, the element-wise explicit CSI of the inter-IRS link still cannot be obtained. For passive IRS, it becomes an even more challenging problem to estimate the double-reflection channel, due to the higher-order cascaded channel coefficients as well as its intricate coupling with the single-reflection channels, which thus calls for designing new channel estimation techniques. 
 Based on the estimated CSI, the active beamforming at the BS and passive beamforming at all IRSs can be jointly designed\cite{Zheng2020DoubleIRS}. {\color{black}Specifically,  it has been shown that deploying two cooperative IRSs near the BS and users achieves superior performance to the case with single-IRS deployment in terms of the maximum SNR  and multi-user effective channel rank\cite{Zheng2020DoubleIRS}. 
However,  robust IRS beamforming for the hybrid IRS deployment under  CSI errors is worth further investigating, which shall greatly improve the performance reliability\cite{zhou2020framework}.} 

% techniques can be developed to improve users' QoS, which is still under-explored. 

Alternatively, the beam-training based method does not require the CSI explicitly. Instead, it is a codebook-based beamforming method that directly searches for the best beam in the spatial domain that yields the optimum performance, which is thus of lower complexity and more scalable with the increasing number of IRS reflecting elements.
% in the network as compared to the channel-estimation based methods {\color{black}{due to the finite number of beam codewords}}.
  While an efficient IRS beam training scheme has been recently proposed in\cite{wang2021fast,you2020fast} that is  applicable to systems under the user-/BS-side  IRS deployment, its extension to the hybrid IRS deployment is more challenging due to the additional inter-IRS channels. {\color{black}Specifically, for IRS-aided communication systems in high-frequency bands (e.g., millimeter wave),  the involved channels can be characterized by a small number of propagation paths. In this case, an efficient approach is by designing the hybrid offline-online beam training\cite{mei2021distributed}, where the beam training for the (sparse) quasi-static BS-IRS and inter-IRS channels are performed offline prior to data transmissions, while the channel path directions from each IRS to the mobile users are trained in real time.}  In practice, the beam training method can be used to generate a proper fingerprinting database that collects a set of optimal beamforming designs at some user positions, with which  
deep learning techniques can be leveraged to predict the mapping between the position of an arbitrary user and its corresponding optimal beamforming. More research along this direction is needed for further exploration.

\subsection{{IRS Placement and Elements Allocation}}

While IRS deployment is assumed to be  fixed in the preceding subsections, we consider in this subsection the  design of IRS deployment, including the IRSs' placement and their reflecting elements' allocation.

First, for each user-/BS-side  IRS,  there are generally two placement strategies, namely, the \emph{centralized placement} where all its reflecting elements are placed on one single surface, versus the \emph{distributed placement} where the available reflecting elements  form multiple smaller-size sub-IRSs that are placed  at different locations around the BS/users. Compared with centralized IRS placement, distributed IRS placement in general achieves better 
% provides more flexibility in improving 
 channel conditions, e.g., increasing the LoS probabilities of the IRS-user and/or the BS-IRS  links, and improving their  channel ranks by exploiting the spatial path diversity.
   However, distributed IRS placement generally requires higher signaling overhead and hardware cost as more IRS controllers need to be  installed, each for one sub-IRS.

Next, under  the hybrid IRS deployment with centralized/distributed IRS placement, a unique design challenge lies in how to optimally allocate the  available reflecting elements  between the BS- and user-side IRSs for optimizing the network performance {\color{black}given a fixed number of total reflecting elements}.  {\color{black}To draw useful insights into this elements-allocation problem, we first consider two simplified system setups.} First, for a multi-user system where  all reflecting elements are assumed to be deployed  near either  the BS or distributed users, it has been shown in\cite{zhang2020intelligentcapa} that the BS-side IRS deployment yields superior rate performance to the user-side deployment under a ``twin" channel setup (i.e., symmetric  channel realizations for fair comparison). Second, for a single-user system,
 it was revealed in\cite{Zheng2020DoubleIRS} that, the available reflecting elements should be split to form two comparable-size IRSs, each deployed near the BS and the user, respectively, for reaping the double-reflection multiplicative beamforming gain.
 %deploying two cooperative IRSs near the BS and user is shown to achieve higher rate than placing all reflecting elements at either BS/user. 
{\color{black}The above  two results indicate that for rate maximization under the hybrid IRS deployment, the available reflecting elements should be properly allocated to both the BS- and user-side IRSs. However, under the proposed hybrid IRS deployment with generally multiple users and time-varying  channel conditions (e.g., the LoS availability) in the links between any two nodes,
how to solve this general IRS elements allocation problem is still open. A viable approach to solve this problem is by firstly determining the IRS-user association based on the statistical CSI and then optimizing the IRS elements allocation by using the  continuous variable  relaxation and integer-rounding techniques for e.g., maximizing the minimum rate among all users. Moreover, for a large-scale network with massive randomly distributed users, the information-theoretical approach adopted in \cite{Zheng2020DoubleIRS} and \cite{zhang2020intelligentcapa}  for optimizing IRS elements  allocation might be computationally formidable. As such, new mathematical tools such as stochastic geometry and deep learning may need to be invoked to facilitate the network-level analysis and efficient elements allocation.

\begin{figure}[t]
	\centering
	\includegraphics[width=3.5in]{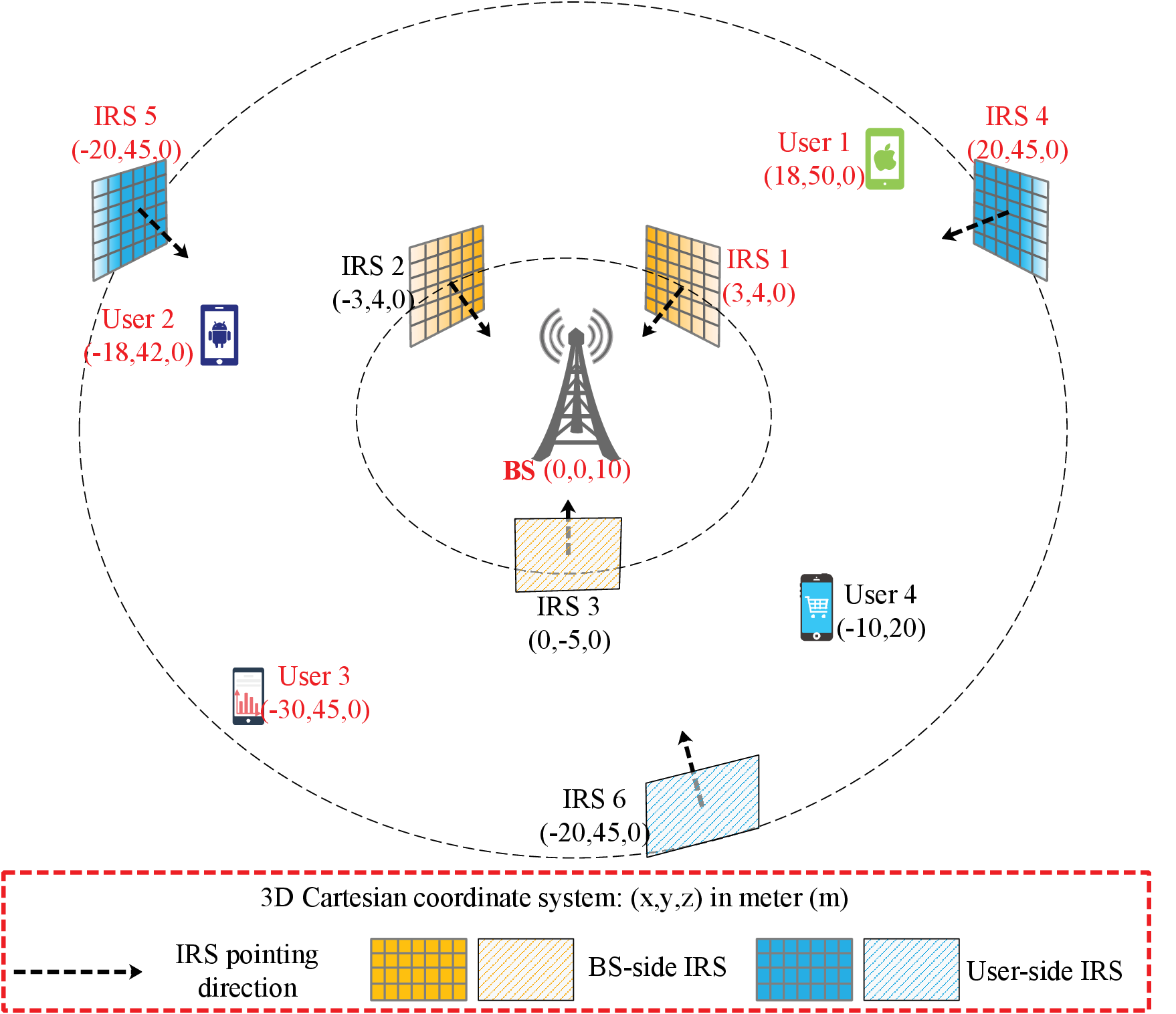}
	\setlength{\abovecaptionskip}{5pt}
	\caption{Simulation setup under the 3D Cartesian coordinate system with $(x,y,z)$ in m.}
	\label{Simulation}
	%\vspace{-0.8cm}
\end{figure}
\section{Simulation Results}\label{Sim}
In this section, we present simulation results to demonstrate the effectiveness of the proposed hybrid IRS deployment.
{\color{black}We consider a typical IRS-aided single-cell system under the general hybrid IRS deployment as shown in Fig.~\ref{Simulation}.
Specifically, three BS-side IRSs (labeled as IRSs 1, 2, and 3) and three user-side IRSs (labeled as IRSs 4, 5, and 6) are deployed
to assist in the uplink transmissions from four single-antenna
users to a 20-antenna BS, with their locations specified in Fig.~\ref{Simulation}. 
Moreover, the half-space reflection of each IRS is specified by the IRS pointing direction.
%, as shown in Fig.~\ref{Simulation}.
For example, (BS-side) IRS 1 can serve the cell-edge users 2, 3, and 4 in its serving half-space, while user 1 is outside its serving space. 
%On the other hand, (user-side) IRS 4 can serve user 1 only within its local coverage.
In addition, we assume that the links between the BS/users and their nearby serving IRSs follow the LoS channel model with the path-loss exponent of $2.2$; the links between the BS/users and their far-apart IRSs are modeled by the LoS-dominant Rician
fading channel with the Rician factor of $5$ dB and path-loss exponent of $2.5$; and the BS-user links follow the Rayleigh fading channel model with the path-loss exponent of $3$. 
The reference channel power gain at the distance of 1 meter (m) is $-30$~dB.
All the users are assumed to have equal transmit power of $5$ dBm
and the noise power at the BS receiver is $-65$ dBm.
Moreover, the multi-antenna BS applies the maximal-ratio combining (MRC) receive beamforming to the received signal from each user.}

\begin{figure}[t]
	\centering
	\includegraphics[width=3in ]{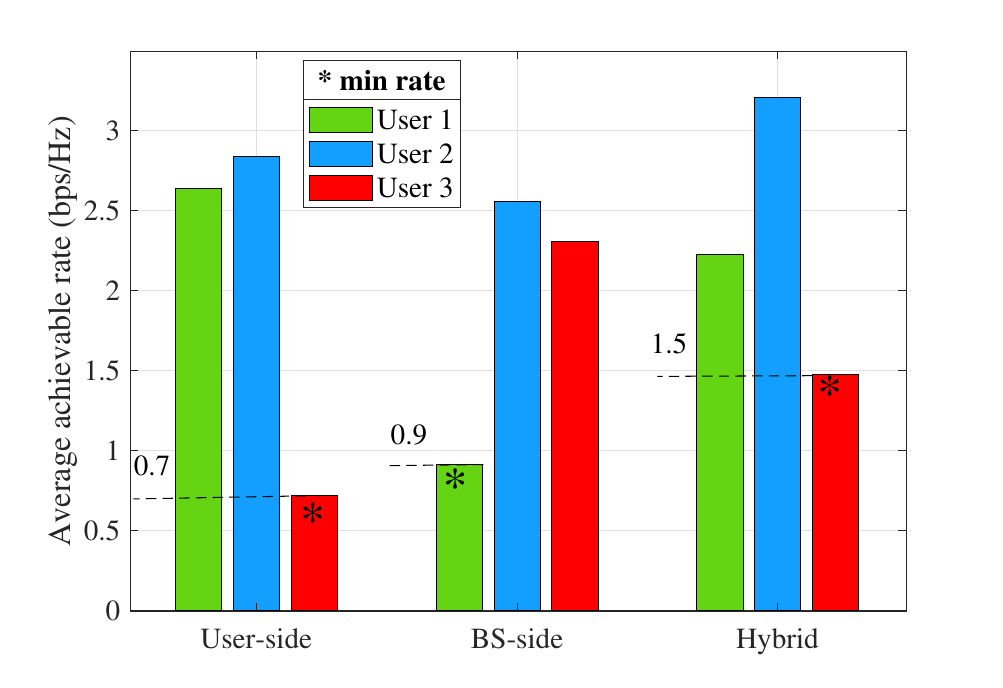}
	\setlength{\abovecaptionskip}{-5pt}
	\caption{Achievable rates of the three typical users as well as their
		min-rate under different IRS deployment strategies.}
	\label{hybridIRS_compV3}
	%\vspace{-0.5cm}
\end{figure}
\begin{figure}[!t]
	\centering
	\includegraphics[width=3in ]{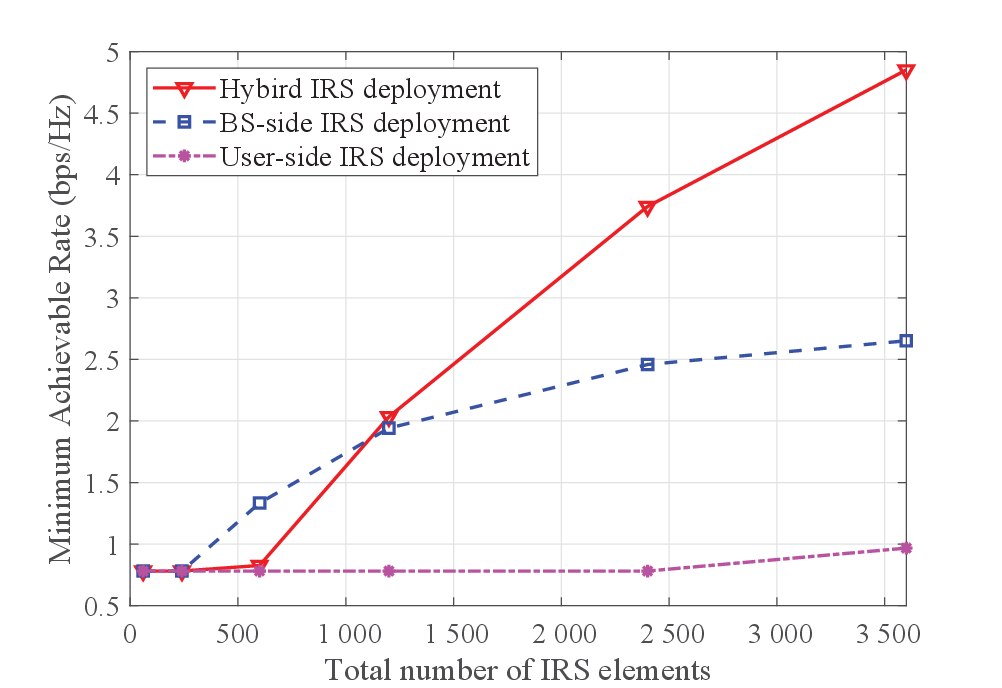}
	\caption{{\color{black}Achievable min-rate versus total number of passive reflecting elements, $M$.}}
	\label{IRSdeploy_element}
	%\vspace{-0.5cm}
\end{figure}

{\color{black}
To showcase the effectiveness of the hybrid IRS deployment, we consider a typical case where IRSs $1$, $4$, and $5$ are employed to assist the communications of users $1$, $2$, and $3$ under the three IRS deployment strategies, while the other IRS randomly tunes their reflections and thus are treated as environmental scatterers.} 
Given a total number of $600$ reflecting elements, we consider the following elements allocation for the three
IRS deployment strategies under the considered system setup:
1) IRSs $4$ and $5$ are equipped with $300$ reflecting elements for the user-side IRS deployment (only); 
2) IRS $1$ is equipped with $600$ reflecting elements for the BS-side IRS deployment (only);
and 3) IRSs $1$, $4$, and $5$ are equipped with $300$, $150$, and $150$ reflecting elements, respectively, for the hybrid IRS deployment.  Moreover, for (BS-side) IRS $1$, we apply the exhaustive search to find its optimal surface partition for maximizing its minimum achievable rate (min-rate) among all users.
{\color{black}In Fig. \ref{hybridIRS_compV3}, we compare the min-rate in bits per second per Hertz (bps/Hz) under the three IRS deployment strategies.}  First, it is observed that under the user-side IRS deployment, user $3$
achieves a much lower rate than users $1$ and $2$, thus becoming the min-rate performance bottleneck in the system, as the cell-edge user $3$ is outside the local coverage of any user-side IRS.
% and thus served by the
%BS directly. 
Next, the BS-side IRS deployment strategy is observed to effectively enhance the achievable rate of user $3$
without sacrificing much the rate performance of user $2$, thanks to its great flexibility in the surface partition and
passive beamforming. In this case, user $1$ becomes the min-rate performance bottleneck since it is not in the serving half-space of the BS-side IRS (i.e., IRS $1$).
% and thus cannot enjoy the passive beamforming gain. 
 Third, with proper reflecting elements allocation as well as efficient BS-side IRS surface partition, the hybrid IRS deployment strategy is observed to achieve a much higher min-rate than both the user-side and BS-side IRS deployment strategies, since 
 % due to the following
%reasons. First, in the considered setup, 
all the users can be covered by at least one (BS-side or user-side) IRS under the
hybrid IRS deployment.
%which demonstrates its more effectiveness in terms of network coverage. Second, 
Moreover, user $2$ has the
potential to enjoy a higher-order passive beamforming gain from the double-reflection link (as compared to the single-reflection link), and hence more reflecting elements at the BS-side IRS can be assigned to associate with user $3$ for improving its achievable rate over the single-reflection link only, thus more flexibly balancing their rates for improving the system min-rate.

{\color{black}Next,  
we show in Fig. \ref{IRSdeploy_element} the min-rate versus the total number of reflecting elements (denoted by $M$) for different IRS deployment strategies, where all the nodes shown in Fig.~\ref{Simulation} are considered.
For each IRS deployment strategy, we apply the exhaustive search to find its optimal user-IRS association to maximize its min-rate. Moreover, we consider the following elements allocation for the three
IRS deployment strategies: 1) each user-side IRS is equipped with $M/3$ reflecting elements for the user-side IRS deployment (only); 
2) each BS-side IRS is equipped with $M/3$ reflecting elements for the BS-side IRS deployment (only);
and 3) each IRS is equipped with $M/6$ reflecting elements for the hybrid IRS deployment.
It is observed that as $M$ increases, the min-rates of all IRS deployment strategies increase thanks to the more pronounced passive beamforming gain from  IRSs.
However, the min-rate improvement of the user-side IRS deployment is marginal when $M$ increases, since some users are outside the user-side IRSs' local coverage. In contrast, 
%although also limited by the number of available IRSs, 
the BS-side IRS deployment strategy achieves significant rate improvement as $M$ increases, since the BS-side IRSs can potentially cover all users in their reflection half-space.
% by exploiting its  much larger serving coverage.
Moreover, when $M<1~000$, the BS-side IRS deployment strategy achieves better performance than the hybrid counterpart, since the single-reflection links tend to provide higher performance gains than the double-reflection link when $M$ is small.
% and the number of reflecting elements per IRS in the BS-side IRS deployment is twice that in the hybrid IRS deployment.
However, when $M$ is sufficiently large (i.e., $M>1~200$), the hybrid IRS deployment strategy outperforms the other two IRS deployment strategies, 
%This is expected 
because it can achieve their complementary advantages and exploit the more pronounced inter-IRS cooperative passive beamforming gain with a large $M$.
%with more flexible user-IRS association, thus effectively balancing users' rates for improving the min-rate.
}

\section{Conclusions}
In this article, we provide an overview of the IRS deployment design in wireless networks from a communication perspective. We first 
%discuss the potential benefits of 
compare  the conventional   BS- versus  user-side IRS deployment strategies, and then introduce a new   hybrid IRS deployment strategy by combining their complementary advantages. 
% with IRSs deployed at both sides to achieve enhanced network performance.
%  hybrid IRS deployment  are first discussed, as compared to the BS- and user-side deployment. 
The main challenges for designing the IRS-aided wireless network under the general hybrid deployment are discussed, including the IRS-user association and mode selection, IRS channel acquisition and beamforming optimization, and IRS placement and elements allocation. Promising methods to tackle these challenges as well as directions for future investigation are provided.   % , including IRS association and mode selection, CSI acquisition and beamforming optimization, as well as IRS placement and allocation. 
Numerical results are also presented  to corroborate our discussions and show the performance gains of the hybrid IRS deployment as compared to its one-sided counterparts. 
 
In future work, it is also interesting to study 
%that are worth pursuing 
the more general multi-IRS aided communication system with multi-hop (i.e., more than two hops) signal reflection\cite{mei2021int}. In this case, IRS beamforming needs to be jointly designed with data routing over multiple selected IRSs for serving each user\cite{mei2020cooperative,mei2021mbmh}. 
{\color{black} Besides, it is worth investigating efficient IRS deployment designs tailored for different IRS applications and systems (e.g., relaying system\cite{kang2021irs}, millimeter wave systems\cite{ntontin2020ris}, NOMA systems\cite{mu2021joint}, full-duplex communication\cite{cai2022dep}, vehicle-to-everything (V2X) communication\cite{ozcan2021ris}, radar communication\cite{lu2021intelligent}), as well as various IRS architectures (e.g., aerial IRS\cite{lu2021aerial}, active IRS\cite{you2022act}, relaying IRS\cite{zheng2021relaying}, intelligent refracting/transmitting surface (IRS/ITS)\cite{liu2021star}).
Moreover, in this article, we mainly focus on the IRS deployment design from the signal processing and optimization perspective, while other useful tools such as stochastic geometry\cite{kishk2020exploiting} and machine learning\cite{liu2020ris} can also be applied to efficient IRS deployment designs, especially for large-scale wireless networks, which deserve further investigation.}

\end{document}